\documentclass[aps,prl,10pt,twocolumn,superscriptaddress,showpacs,showkeys]{revtex4-1} 

\def\be{\begin{equation}}
\def\ee{\end{equation}}

\def\C{^\circ \mathrm{C}}
\usepackage{float}
\usepackage[pdftex]{graphicx}
\usepackage{setspace}
\usepackage{amsmath}
\usepackage{amssymb}
\usepackage{xcolor}
\usepackage{siunitx}
\usepackage{dcolumn}
\usepackage{bm}
\usepackage{hyperref}
\usepackage{soul}

\usepackage{caption} 
\renewcommand{\paragraph}[1]{\vspace{0.1cm}\emph{#1} ~--~}

\begin{document}
\title{Soft Condensation}

\author{Ambre Bouillant} 
\affiliation{Physics of Fluids Group, University of Twente, 7500 AE Enschede, Netherlands}
\affiliation{MSC, UMR 7057 CNRS, UPC, 10 rue A. Domon et L. Duquet, 75013, Paris, France}
  \author{Christopher Henkel}
 \affiliation{Institut f\"ur Theoretische Physik, Universit\"at M\"unster, W.Klemm-Str.\ 9, 48149 M\"unster, Germany}
  \author{Uwe Thiele}
 \affiliation{Institut f\"ur Theoretische Physik, Universit\"at M\"unster, W.Klemm-Str.\ 9, 48149 M\"unster, Germany}
%
 \author{Bruno Andreotti}
 \affiliation{LPENS, UMR 8550 CNRS, ENS, UPCité, SU, 24 rue Lhomond, 75005 Paris, France}
%
\author{ Jacco H. Snoeijer}
\affiliation{Physics of Fluids Group, University of Twente, 7500 AE Enschede, Netherlands}

\date{\today}

\begin{abstract}
When moist air meets a cold surface, it creates a breath figure characterized by numerous small droplets. The central question is how the vapor flux is distributed between the growth of previously condensed drops and the nucleation of new ones. Here, we investigate the nucleation, growth, and coalescence of droplets on soft crosslinked polymer networks. The number of droplets initially remains constant, until drops start to coarsen according to a universal law; both phenomena are explained via the formation of a saturated boundary layer. Although nucleation occurs at a scale where the polymer network resembles a melt, we quantitatively unveil an algebraic sensitivity of the number of droplets on the substrate elasticity. Our findings suggest that nucleation follows a surprisingly low-energy pathway, influenced by the degree of crosslinking. Consequently, breath figures offer a macroscopic approach to probe the molecular characteristics of the polymer interface.
\end{abstract}

\maketitle

When humid air is directed onto a mirror, a fine mist of water droplets condenses on its surface. The condensation involves the emergence of nascent structures of a new phase, a phenomenon known as nucleation. Nucleation is prevalent in nature, and is fundamental across various domains including cloud condensation, cell biology, material sciences, food industry and water collection \cite{kelton2010nucleation,kashchiev2000nucleation}.
The patterns formed by dew on a substrate, known as ``breath figures" \cite{Rayleigh1911}, have been thoroughly investigated on various substrate types \cite{Knobler1991},  from rough \cite{Park2016,Trosseille2019}, chemically heterogeneous 
or patterned \cite{Varanasi2009,Bintein2019}, polymeric \cite{Briscoe1991,Katselas2022,Leach2006,Phadnis2017,Sokuler2010,Stricker2022universality}, liquid-infused \cite{Anand2015,Sharma2022} to pure liquid substrates \cite{Steyer1990,Steyer1993}.
The collected droplets typically evolve towards scale-free size distributions, ranging from freshly nucleated nanometer-scale drops to millimeter sizes after growth and coalescence \cite{Stricker2022universality,beysens1986growth,Viovy1988scaling,family1988scaling,Kolb1989,blaschke2012breath,brilliantov1998polydisperse,family1989kinetics}. 
A remarkable unexplained feature is that on atomically smooth substrates, the droplet patterns initially appear to be nearly monodisperse \cite{Steyer1990,Steyer1993}, as if all drops appeared simultaneously through a single nucleation burst.

In the context of surface patterning and self-assembly
\cite{Guha2017,Baratian2018,boker2004hierarchical,zhang2015breath}, highly deformable reticulated polymer networks have shown to offer new routes to engineer surface properties \cite{style2013patterning,Sokuler2010,dent2022temporally}. Recent studies have resolved the long-standing question of how sessile, isolated, macroscopic drops wet soft elastic surfaces \cite{Andreotti2020Annual}. 
The equilibrium shape of isolated sessile macroscopic drop is dictated by the so-called elastocapillary length $\gamma/G$, which compares the surface tension $\gamma$ with the substrate elastic shear modulus $G$. Droplets smaller than $\gamma/G$ sink into the substrate to form liquid lenses and can interact with nearby drops via the inverted ``Cheerios effect" \cite{Karpitschka2016}. Whether the classical elastocapillary framework accurately describes the key features of soft condensation, from the nucleation of isolated nanodroplets to collective effects, remains an open question that we address here. 

In this Letter, we experimentally investigate the condensation of water vapor on soft polymeric substrates with elastic moduli $G\in \left[10^2 ; 10^6\right] \;{\rm Pa}$. Figure~\ref{fig:Setup} shows that the number of droplets that condense varies dramatically upon changing the substrate stiffness. 
This observation can not at all be explained by the elastocapillary framework, which predicts nanometric nuclei to be insensitive to the degree of reticulation, since their size is much smaller than $\gamma/G$. A central result is that, in contrast to previous studies, the distribution of drop sizes is nearly monodisperse and the nucleation has completely stopped once drops reach the optically accessible range of a few microns. We develop a quantitative model that explains the arrest of nucleation, and the subsequent growth and coarsening, offering a route towards the actual rate of nucleation at the nanoscale.
\begin{widetext}
\noindent
\begin{minipage}{\textwidth}
\centering
\vspace{-5mm} \includegraphics[width=1\textwidth]{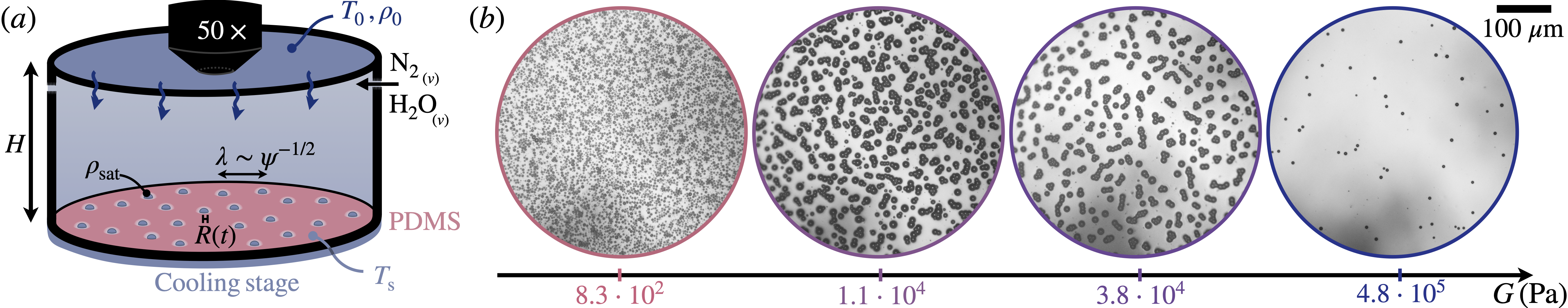}
\end{minipage}
\vspace{-2.5mm} \captionof{figure}{\label{fig:Setup} (a) Schematic of the controlled humidity chamber.  (b) Breath figures imaged from above using a microscope, on substrates of increasing elastic modulus $G$ for $\rho/\rho_{\rm sat}(T_{\rm s})=1.22\pm 0.11$.}
\end{widetext}

\paragraph{Experimental set-up} Materials and methods are detailed in the Supplementary Materials (SM). We prepare millimeter-thick soft reticulated polymer networks by premixing and curing silicone (Dow Corning; Sylgard 184) with its reticulant agent in weight proportions ranging from 10:1 up to 80:1 
or two prepolymer components (DowCorning; CY52-276 in proportions 1.3:1 and 1:1 \& Zhermack; PVS Elite Double 8 elastomer).
Tuning the ratio of base polymer to cross-linker, the elastic modulus $G\in\left[10^2 - 10^6\right] \;{\rm Pa}$, while keeping the same chemical composition. It should be noted that several properties, such as the density of dangling ends and the residual melt inside the network, vary simultaneously.  In the remainder, we use $G$ as the label for the tested substrates. The results presented here are similar whether uncrosslinked polymer chains are extracted or not (cf.  SM). Experiments are also performed on a stiff PDMS pseudo-brush (cf. SM); drops cannot sink into these substrates and thus form spherical caps of contact angle $\theta_{\rm brush}= 100\pm 8^{\circ}$. The low $G$ limit is probed by experiments on uncrosslinked PDMS melt. Macroscopic drops on all tested substrates exhibit no contact angle hysteresis or pinning. Defects are ruled out as drops do not form at the same locations if we run successive experiments.

The condensation is driven by supersaturation that leads to a difference in chemical potential $\Delta \mu=-k_B T\ln\left( \rho   \big/   \rho_{\mathrm{sat}} \right)$ between the  liquid and vapor phase, where $\rho$ is the water vapor mass per unit volume and $\rho_{\rm sat}$ its saturation value. Experiments are performed in a closed cell of height $H=42\rm \;mm$ equipped with a controlled humid/dry air inlet that maintains humidity $\rho=\rho_0$ constant at the ceiling, maintained at temperature $T_0= 22\pm 4\C$ [Fig.~\ref{fig:Setup}($a$)]. The substrate is placed on a cooling Peltier stage at the bottom of the chamber. Quenching the substrate temperature to $T_{\rm s}=5\pm 1\C$ in a few seconds triggers the nucleation and growth of nanometric dew drops on the substrate. The breath figures are illuminated and recorded from above with a Nikon camera (D850) mounted on an upright Olympus microscope with $\times 50$ magnification. This yields a spatial resolution of $12\;{\rm pixels/\mu m}$, allowing for the detection of droplets once their size reaches the micron scale. 
\begin{figure}
\includegraphics[width=.9\columnwidth]{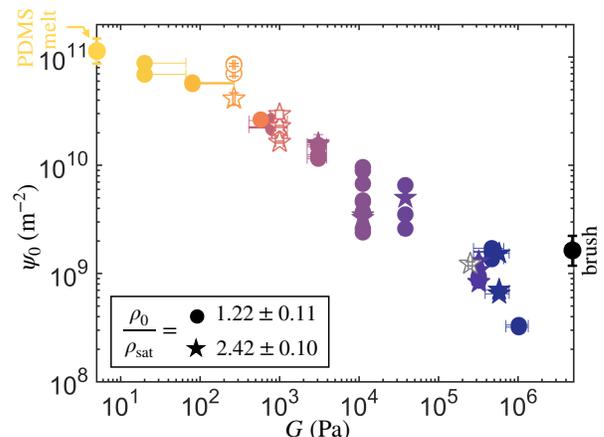}
\vspace{-3mm}
\caption{\label{fig:Setup_b} Drop initial surface density $\psi_0$ measured once drops become visible ($R \sim 1 \mathrm{\mu m}$) as a function of $G$.  Values for PDMS melt and stiff brush are also reported.
\vspace{-9mm}}
\end{figure}

\paragraph{Nucleation} Figure~\ref{fig:Setup}($b$) shows typical drop patterns.  Changing the degree of reticulation of the polymer network dramatically affects the density of the drops: softer substrates clearly favor nucleation. In stark contrast with \cite{Stricker2022universality,beysens1986growth,Viovy1988scaling,family1988scaling,blaschke2012breath,brilliantov1998polydisperse,family1989kinetics}, the distribution of drop sizes is nearly monodisperse and almost no new drops form, at least after they reach the visible micron scale. During a time $\tau_{\rm p}$ ranging from $10^2-10^3 \mathrm{s}$, the number of drops per unit area $\psi_0$ remains constant, while the drops increase in volume. Figure~\ref{fig:Setup_b} reports the droplet surface density $\psi_0$ versus $G$. The values of $\psi_0$ appear to exhibit an algebraic dependence on $G$, and approximately fall between the two limiting cases of the uncrosslinked melt (yellow symbol) and the stiff brush (black symbol). 

The key questions to address are why the number of droplets varies so strongly with softness, and why hardly any new drops are seen to nucleate during the experiment --leading to quasi-monodisperse patterns. The latter observation is striking since nucleation is stochastic, with a nucleation rate $J$ per unit surface that increases with the degree of supersaturation. The number of nucleated drops thus increases as $\psi \sim J t$ during the initial stages of the experiment, at scales below optical resolution. Below we show that interactions between neighbouring drops lead to the arrest of nucleation, and to the selection of $\psi_0$, in the experimentally observable regime. 

\paragraph{Growth and arrest of nucleation } The typical distance between drops $\psi^{-1/2}$ is initially much larger than their average radius $R$ [Fig.~\ref{fig:Setup}($a$)]. Therefore, the drops can be considered isolated and grow according to
\begin{equation}\label{eq:isolated}
R^2 \sim  D (\rho_0 - \rho_{\rm sat}) t/ \rho_\ell,
\end{equation}
where $D=2.3 \cdot 10^{-5} \,{\rm m^2/s}$ is the diffusion constant of water vapor in air, and  $\rho_{\ell}=10^3 \rm \; kg/m^3$. As  drops grow, collective effects slow down this condensation dynamics \cite{Sokuler2010interspacing,Viovy1988scaling}. Indeed, droplets are saturated at their liquid-vapor interface and thereby decrease the local humidity around them [Fig.~\ref{fig:Setup}($a$)]. We show in a companion paper using a matched asymptotics description of the diffusion problem \cite{paper_bis} that the drops gradually build up a boundary layer: seen from the scale of the cell, the substrate becomes saturated at an effective humidity $\rho_{\rm eff}$ \cite{paper_bis},  
\begin{equation}\label{eq:phieff}
\rho_{\rm eff} = \frac{\rho_0 + 2\pi \mathcal F \psi R H \rho_{\rm sat} }{1+2\pi \mathcal F \psi R H},
\end{equation}
where $\mathcal F$ is a known factor that accounts for the geometry of the emerged drop \cite{Popov2005}. 

Equation (\ref{eq:phieff}) predicts two distinct regimes. In the early stage of the condensation process, that is, in the low density limit, $\psi R H \ll 1$, the effective humidity $\rho_{\rm eff} \approx \rho_0$. 
Dew droplets thus initially grow according to (\ref{eq:isolated}). Conversely, in the high density limit, $\psi R H \gg 1$, the effective humidity at the substrate is well below $\rho_0$ and approaches $\rho_{\rm sat}$. Since $\psi R H\sim 10^2$ when droplets become first visible, our experiments are in this high-density regime. Hence $\rho_{\rm eff} \approx \rho_{\rm sat}$ near the substrate, which explains why the nucleation of new droplets is arrested. The time of arrest $\tau_{\rm a}$ is given by the crossover between the two regimes and can be determined from $\psi \sim Jt$ combined with (\ref{eq:isolated}-\ref{eq:phieff}). This selects the number of drops
\begin{equation}\label{eq:prediction}
\psi_0 \sim J \tau_{\rm a} \sim \left( \frac{J}{H^2 D(\rho_0-\rho_{\rm sat})} \right)^{1/3},
\end{equation}
as is validated by stochastic numerics in \cite{paper_bis}. This scenario explains why in experiments, for which  $t \gg \tau_{\rm a}$, it appears as if all drops nucleated quasi-simultaneously.
\begin{figure}
\includegraphics[width=0.9\columnwidth]{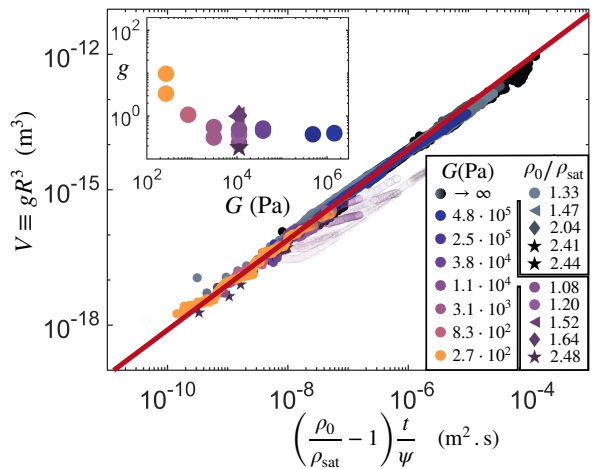}
\vspace{-2mm}
\caption{\label{fig:Growth} ($a$)~Mean drop volume $V\equiv g R^3$ as a function of time rescaled according to Eq.~(\ref{eq:subdiffusivegrowth}) (red line) obtained on brush and soft substrates for various relative humidities $\rho_0/\rho_{\rm sat}$ and moduli $G$. For brushes the geometrical factor $g= 2.6 \pm 0.4$ is measured independently; for soft gels, its fitted value is reported in the inset as a function of $G$. }
\vspace{-6mm}
\end{figure}

Our saturation scenario can be tested experimentally. The diffusive flux across the cell saturates to $D(\rho_0-\rho_{\rm sat})/H$, and each drop receives a fraction of this flux. The average drop volume $V$ is thus expected to grow as
\begin{equation}
 V =\frac{D \left( \rho_0 - \rho_{\mathrm sat} \right)  }{\rho_\ell H }\frac{t}{\psi}.
 \label{eq:subdiffusivegrowth}
\end{equation}
Figure~\ref{fig:Growth} shows that this relation is obeyed for all substrates. To convert the measured (average) radius $R$ to the mean drop volume, we introduce a geometrical factor $g \equiv V/R^3$. On brush substrates, the drops form spherical caps of constant contact angle and allow for an independent calibration ($g=2.6 \pm 0.4$, cf.  SM). Data then follow (\ref{eq:subdiffusivegrowth}) without any adjustable parameter. On soft gels, the drops partly sink into the substrate and exhibit a slow relaxation dynamics, both affecting $g$ in a nontrivial manner. Still, the scaling $R^3 \sim t/\psi$ is seen to hold. The fitted value of $g$ is found to decrease with $G$ when $G< 3\cdot 10^3$~Pa, reaching a plateau when $G> 3\cdot 10^3$~Pa (Fig.~\ref{fig:Growth}, inset). At later times, the measured $R^3$ for very soft gels systematically fall below the prediction (Fig.~\ref{fig:Growth}, light points). This trend is in line with drops sinking into the gel, storing part of their volume below the surface.
\begin{figure}
\includegraphics[width=\columnwidth]{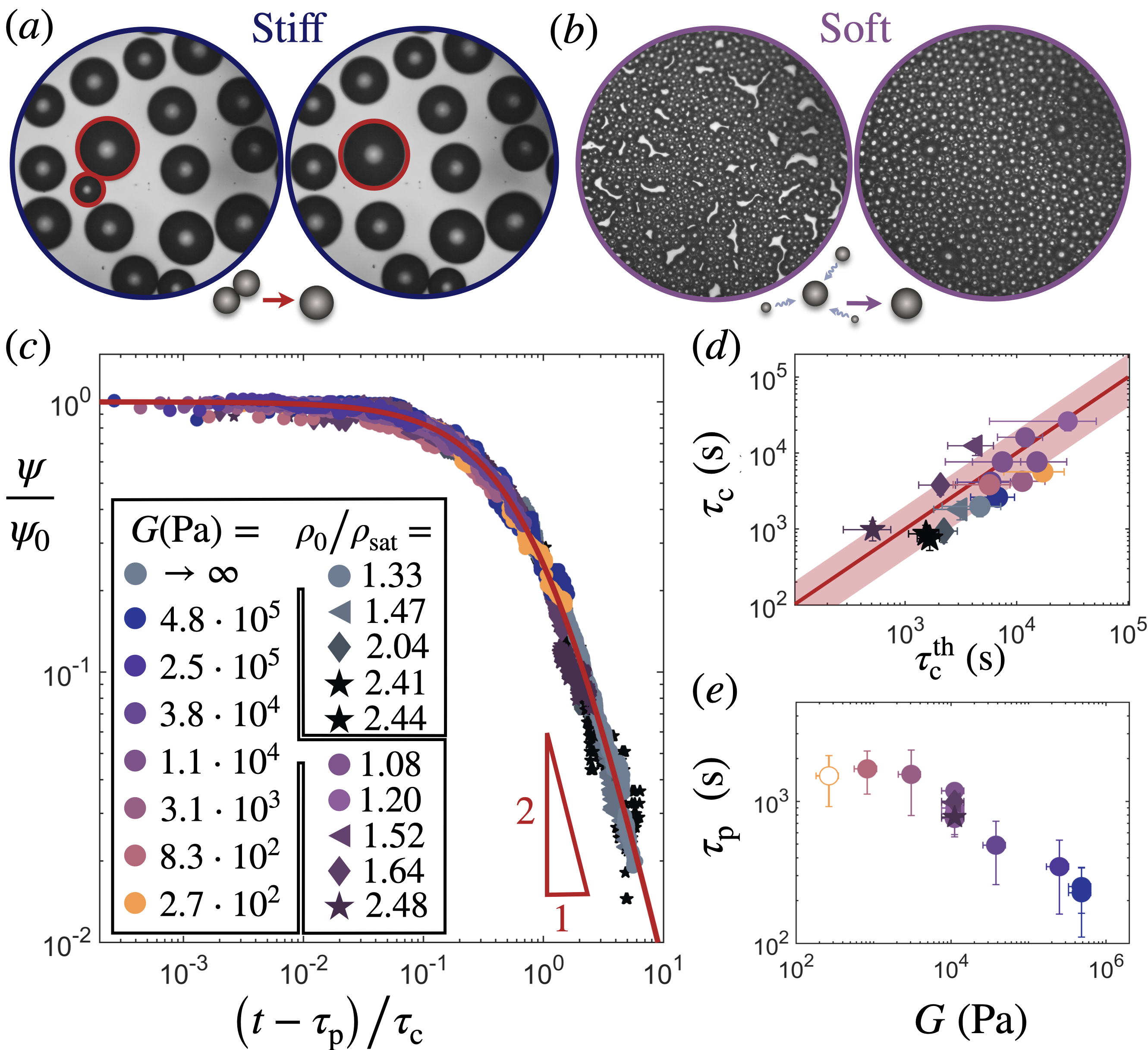}
\caption{\label{fig:Coarsening} Evolution of breath figures (seen from top) on ($a$) brush PDMS (before and after a merging event) ($b$)~and a soft gel ($G=1.1\cdot 10^4$~Pa).
($c$)~Normalized drop surface density $\psi/\psi_0$ for all tested substrates and humidities, as a function of rescaled time $(t-\tau_{\rm p})/\tau_{\rm c}$.  The coarsening time $\tau_{\rm c}$ and delay time $\tau_{\rm p}$ are deduced by fitting (\ref{eq:psieq}). ($d$)~$\tau_{\rm c}$ follows (\ref{eq:psieq}). The red line has a slope~$1 \pm 0.8$. ($e$)~$\tau_{\rm p}$ corresponds to the plateau of $\psi(t)$, plotted as a function of the softness $G$. }
\vspace{-6mm}
\end{figure}

\paragraph{Coarsening} Figure~\ref{fig:Coarsening} shows that after a time $\tau_{\rm p}$, the drop pattern exhibits a universal coarsening where the density $\psi$ decreases in times as $\psi \sim t^{-2}$. We show in the companion paper \cite{paper_bis} that in case of rapid coalescence, the droplet density $\psi(t)$ decreases according to
\begin{equation}
\psi=\frac{\psi_0}{(1+(t-\tau_{\rm p})/\tau_{\rm c})^2}, \quad\tau_{\rm c}^{\rm th} \sim \frac{ \rho_{\ell} H }{ \psi_0^{1/2}D (\rho_0 -  \rho_{\rm sat})},
\label{eq:psieq}
\end{equation}
which is the red line in Fig.~\ref{fig:Coarsening}($c$). On stiff substrates, the coalescence between two contacting drops is nearly instantaneous ($\tau_{\rm p}\approx0$) [Fig.~\ref{fig:Coarsening}($a$)]. On soft substrates, however, droplets display a foam-like structure that resists coalescence [Fig.~\ref{fig:Coarsening}($b$)] and exhibits an Ostwald-like ripening process.  Still, the evolution of $\psi(t)$ is observed to follow the same law, albeit with a significant delay time $\tau_{\rm p}$ before coarsening starts  [Fig.~\ref{fig:Coarsening}($e$)]. The coarsening time $\tau_{\rm c}$ agrees well with prediction (\ref{eq:psieq}), for all tested substrates and humidities [Fig.~\ref{fig:Coarsening}($d$)]. 
\begin{figure}
\includegraphics[width=\columnwidth]{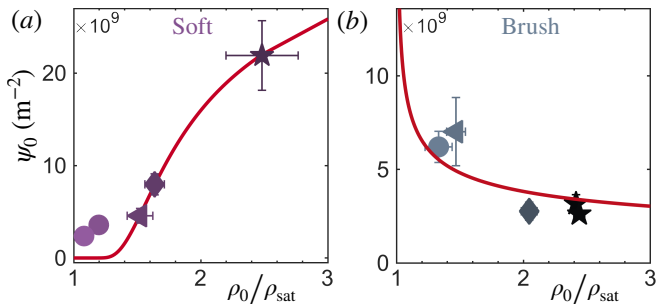}
\vspace{-6mm}
\caption{\label{fig:humidity}  Drop density $\psi_0$ as a function of $\rho_0/\rho_{\rm sat}$ measured on ($a$)~soft PDMS gels ($G=1.1\cdot 10^4$~Pa) and on ($b$)~PDMS brushes. Red lines are the best fits by CNT, reached for $\chi=1.3$ and $\chi=0$ respectively.
\vspace{-7mm}}
\end{figure}

\paragraph{Discussion} In summary, we have shown that the random nucleation of drops ceases when a saturated diffusive boundary layer forms above the substrate and that their coarsening obeys a universal law. The breath figures exhibit a nearly monodisperse distribution even at a later stage, which contrasts with most of the literature on breath figures.
Our study reveals that achieving monodisperse patterns requires: (i) smooth, defect-free surfaces; (ii) operating within the diffusion-controlled regime, ensuring the competition between growth and nucleation regimes to arrest nucleation events and set $\psi_0$ ($t\gg \tau_a$); (iii) the inhibition of coalescence. Notably, forced or natural convection, induced by strong air injection or unstable air stratification when the substrate orientation is vertical or downward, can impede the formation of the diffusive boundary layer near the substrate, thus hindering the selection of $\psi_0$. This may elucidate the observed polydispersity when condensation occurs at vertical substrates \cite{Phadnis2017,Rose2002} or below horizontal subtrates  \cite{Sikarwar2012,Leach2006,family1989kinetics,Kolb1989}. Additionally, the occurrence of coalescence leads to bimodal distribution in droplet sizes \cite{Stricker2022universality}. 
Nevertheless, the patterns obtained at the early stage of the process \cite{Zhao1995,Katselas2022,Sokuler2010} or on substrates able to delay or prevent droplet merging \cite{Guha2017,Briscoe1991,Anand2015,Steyer1990,Steyer1993,Nepomnyashchy2006,Knobler1988,Zhang2020}, are rather monodisperse. 

The main unexplained observation of the experiments is that  the breath figures are much denser when the substrate is softer. As in many other nucleation problems \cite{kashchiev2000nucleation,Merikanto2007,li2021understanding}, the observed number of drops is not at all compatible with classical nucleation theory (CNT). The nucleation rate in CNT is treated as an activated process $J =J_0 \exp\left(-\Delta \mathcal G^*/k_B T\right)$, with an energy barrier $\Delta \mathcal G^* =\chi k_B T/\ln^2\left( \rho   \big/   \rho_{\mathrm{sat}} \right)$. The factor $\chi$ depends on the droplet geometry \cite{paper_bis}. Figure~\ref{fig:humidity}($a$) shows a moderate increase of $\psi_0$ upon increasing relative humidity above a soft substrate, fitted by $\chi=1.3$; by contrast, the macroscopic theory would predict a violent increase associated with $\chi\sim 50$. The failure of CNT is even more striking for brushes, for which $\psi_0$ is seen to decrease with relative humidity as shown in Fig.~\ref{fig:humidity}($b$). The red line corresponds to (\ref{eq:prediction}) while taking a constant value of $J$ (i.e. $\chi=0$ instead of the expected $\chi=50$). The implication of these results is that nucleation occurs via a microscopic path that has a very low barrier, and which is very sensitive to the degree of reticulation, even though the nuclei are much smaller than the distance between cross-linkers. As such, macroscopic breath figures potentially offer a probe for molecular pathways at polymeric interfaces.

The authors thank J. Scheefhals for supporting experiments, and acknowledge financial support from NWO Vici (No. 680-47-632),  and DFG (Nos. TH781/12 and SN145/1-1 within SPP 2171). 

\bibliography{bibmain}

\providecommand{\noopsort}[1]{}\providecommand{\singleletter}[1]{#1}%
\begin{thebibliography}{44}%
\makeatletter
\providecommand \@ifxundefined [1]{%
 \@ifx{#1\undefined}
}%
\providecommand \@ifnum [1]{%
 \ifnum #1\expandafter \@firstoftwo
 \else \expandafter \@secondoftwo
 \fi
}%
\providecommand \@ifx [1]{%
 \ifx #1\expandafter \@firstoftwo
 \else \expandafter \@secondoftwo
 \fi
}%
\providecommand \natexlab [1]{#1}%
\providecommand \enquote  [1]{``#1''}%
\providecommand \bibnamefont  [1]{#1}%
\providecommand \bibfnamefont [1]{#1}%
\providecommand \citenamefont [1]{#1}%
\providecommand \href@noop [0]{\@secondoftwo}%
\providecommand \href [0]{\begingroup \@sanitize@url \@href}%
\providecommand \@href[1]{\@@startlink{#1}\@@href}%
\providecommand \@@href[1]{\endgroup#1\@@endlink}%
\providecommand \@sanitize@url [0]{\catcode `\\12\catcode `\$12\catcode
  `\&12\catcode `\#12\catcode `\^12\catcode `\_12\catcode `\%12\relax}%
\providecommand \@@startlink[1]{}%
\providecommand \@@endlink[0]{}%
\providecommand \url  [0]{\begingroup\@sanitize@url \@url }%
\providecommand \@url [1]{\endgroup\@href {#1}{\urlprefix }}%
\providecommand \urlprefix  [0]{URL }%
\providecommand \Eprint [0]{\href }%
\providecommand \doibase [0]{http://dx.doi.org/}%
\providecommand \selectlanguage [0]{\@gobble}%
\providecommand \bibinfo  [0]{\@secondoftwo}%
\providecommand \bibfield  [0]{\@secondoftwo}%
\providecommand \translation [1]{[#1]}%
\providecommand \BibitemOpen [0]{}%
\providecommand \bibitemStop [0]{}%
\providecommand \bibitemNoStop [0]{.\EOS\space}%
\providecommand \EOS [0]{\spacefactor3000\relax}%
\providecommand \BibitemShut  [1]{\csname bibitem#1\endcsname}%
\let\auto@bib@innerbib\@empty
\bibitem [{\citenamefont {Kelton}\ and\ \citenamefont
  {Greer}(2010)}]{kelton2010nucleation}%
  \BibitemOpen
  \bibfield  {author} {\bibinfo {author} {\bibfnamefont {K.}~\bibnamefont
  {Kelton}}\ and\ \bibinfo {author} {\bibfnamefont {A.}~\bibnamefont {Greer}},\
  }\href@noop {} {\emph {\bibinfo {title} {Nucleation in condensed matter:
  applications in materials \& biology}}}\ (\bibinfo  {publisher} {Elsevier},\
  \bibinfo {year} {2010})\BibitemShut {NoStop}%
\bibitem [{\citenamefont {Kashchiev}(2000)}]{kashchiev2000nucleation}%
  \BibitemOpen
  \bibfield  {author} {\bibinfo {author} {\bibfnamefont {D.}~\bibnamefont
  {Kashchiev}},\ }\href@noop {} {\emph {\bibinfo {title} {Nucleation}}}\
  (\bibinfo  {publisher} {Elsevier},\ \bibinfo {year} {2000})\BibitemShut
  {NoStop}%
\bibitem [{\citenamefont {Rayleigh}(1911)}]{Rayleigh1911}%
  \BibitemOpen
  \bibfield  {author} {\bibinfo {author} {\bibfnamefont {L.}~\bibnamefont
  {Rayleigh}},\ }\href {\doibase 10.1259/jrs.1911.0067} {\bibfield  {journal}
  {\bibinfo  {journal} {J. R{\"{o}}ntgen Soc.}\ }\textbf {\bibinfo {volume}
  {7}},\ \bibinfo {pages} {126} (\bibinfo {year} {1911})}\BibitemShut {NoStop}%
\bibitem [{\citenamefont {{CM. Knobler}}\ \emph {et~al.}(1991)\citenamefont
  {{CM. Knobler}}, \citenamefont {Steyer}, \citenamefont {Guenoun},\ and\
  \citenamefont {Fritter}}]{Knobler1991}%
  \BibitemOpen
  \bibfield  {author} {\bibinfo {author} {\bibnamefont {{CM. Knobler}}},
  \bibinfo {author} {\bibfnamefont {A.}~\bibnamefont {Steyer}}, \bibinfo
  {author} {\bibfnamefont {P.}~\bibnamefont {Guenoun}}, \ and\ \bibinfo
  {author} {\bibfnamefont {D.}~\bibnamefont {Fritter}},\ }\href {\doibase
  10.1080/01411599108206932} {\bibfield  {journal} {\bibinfo  {journal} {Phase
  Transitions}\ }\textbf {\bibinfo {volume} {31}},\ \bibinfo {pages} {219}
  (\bibinfo {year} {1991})}\BibitemShut {NoStop}%
\bibitem [{\citenamefont {{KC. Park}}\ \emph {et~al.}(2016)\citenamefont {{KC.
  Park}}, \citenamefont {Kim}, \citenamefont {Grinthal}, \citenamefont {He},
  \citenamefont {Fox}, \citenamefont {{JC. Weaver}},\ and\ \citenamefont
  {Aizenberg}}]{Park2016}%
  \BibitemOpen
  \bibfield  {author} {\bibinfo {author} {\bibnamefont {{KC. Park}}}, \bibinfo
  {author} {\bibfnamefont {P.}~\bibnamefont {Kim}}, \bibinfo {author}
  {\bibfnamefont {A.}~\bibnamefont {Grinthal}}, \bibinfo {author}
  {\bibfnamefont {N.}~\bibnamefont {He}}, \bibinfo {author} {\bibfnamefont
  {D.}~\bibnamefont {Fox}}, \bibinfo {author} {\bibnamefont {{JC. Weaver}}}, \
  and\ \bibinfo {author} {\bibfnamefont {J.}~\bibnamefont {Aizenberg}},\ }\href
  {\doibase 10.1038/nature16956} {\bibfield  {journal} {\bibinfo  {journal}
  {Nature}\ }\textbf {\bibinfo {volume} {531}},\ \bibinfo {pages} {78}
  (\bibinfo {year} {2016})}\BibitemShut {NoStop}%
\bibitem [{\citenamefont {Trosseille}\ \emph {et~al.}(2019)\citenamefont
  {Trosseille}, \citenamefont {Mongruel}, \citenamefont {Royon}, \citenamefont
  {{MG. Medici}},\ and\ \citenamefont {Beysens}}]{Trosseille2019}%
  \BibitemOpen
  \bibfield  {author} {\bibinfo {author} {\bibfnamefont {J.}~\bibnamefont
  {Trosseille}}, \bibinfo {author} {\bibfnamefont {A.}~\bibnamefont
  {Mongruel}}, \bibinfo {author} {\bibfnamefont {L.}~\bibnamefont {Royon}},
  \bibinfo {author} {\bibnamefont {{MG. Medici}}}, \ and\ \bibinfo {author}
  {\bibfnamefont {D.}~\bibnamefont {Beysens}},\ }\href@noop {} {\bibfield
  {journal} {\bibinfo  {journal} {Eur. Phys. J. E}\ }\textbf {\bibinfo {volume}
  {42}} (\bibinfo {year} {2019})}\BibitemShut {NoStop}%
\bibitem [{\citenamefont {{KK. Varanasi}}\ \emph {et~al.}(2009)\citenamefont
  {{KK. Varanasi}}, \citenamefont {Hsu}, \citenamefont {Bhate}, \citenamefont
  {Yang},\ and\ \citenamefont {Deng}}]{Varanasi2009}%
  \BibitemOpen
  \bibfield  {author} {\bibinfo {author} {\bibnamefont {{KK. Varanasi}}},
  \bibinfo {author} {\bibfnamefont {M.}~\bibnamefont {Hsu}}, \bibinfo {author}
  {\bibfnamefont {N.}~\bibnamefont {Bhate}}, \bibinfo {author} {\bibfnamefont
  {W.}~\bibnamefont {Yang}}, \ and\ \bibinfo {author} {\bibfnamefont
  {T.}~\bibnamefont {Deng}},\ }\href {\doibase 10.1063/1.3200951} {\bibfield
  {journal} {\bibinfo  {journal} {Appl. Phys. Lett.}\ }\textbf {\bibinfo
  {volume} {95}},\ \bibinfo {pages} {2007} (\bibinfo {year}
  {2009})}\BibitemShut {NoStop}%
\bibitem [{\citenamefont {{PB. Bintein}}\ \emph {et~al.}(2019)\citenamefont
  {{PB. Bintein}}, \citenamefont {Lhuissier}, \citenamefont {Mongruel},
  \citenamefont {Royon},\ and\ \citenamefont {Beysens}}]{Bintein2019}%
  \BibitemOpen
  \bibfield  {author} {\bibinfo {author} {\bibnamefont {{PB. Bintein}}},
  \bibinfo {author} {\bibfnamefont {H.}~\bibnamefont {Lhuissier}}, \bibinfo
  {author} {\bibfnamefont {A.}~\bibnamefont {Mongruel}}, \bibinfo {author}
  {\bibfnamefont {L.}~\bibnamefont {Royon}}, \ and\ \bibinfo {author}
  {\bibfnamefont {D.}~\bibnamefont {Beysens}},\ }\href {\doibase
  10.1103/PhysRevLett.122.098005} {\bibfield  {journal} {\bibinfo  {journal}
  {Phys. Rev. Lett.}\ }\textbf {\bibinfo {volume} {122}},\ \bibinfo {pages}
  {98005} (\bibinfo {year} {2019})}\BibitemShut {NoStop}%
\bibitem [{\citenamefont {{BJ. Briscoe}}\ and\ \citenamefont {{KP.
  Galvin}}(1991)}]{Briscoe1991}%
  \BibitemOpen
  \bibfield  {author} {\bibinfo {author} {\bibnamefont {{BJ. Briscoe}}}\ and\
  \bibinfo {author} {\bibnamefont {{KP. Galvin}}},\ }\href {\doibase
  10.1016/0166-6622(91)80126-9} {\bibfield  {journal} {\bibinfo  {journal}
  {Colloids Surf.}\ }\textbf {\bibinfo {volume} {56}},\ \bibinfo {pages} {263}
  (\bibinfo {year} {1991})}\BibitemShut {NoStop}%
\bibitem [{\citenamefont {Katselas}\ \emph {et~al.}(2022)\citenamefont
  {Katselas}, \citenamefont {Parin},\ and\ \citenamefont
  {Neto}}]{Katselas2022}%
  \BibitemOpen
  \bibfield  {author} {\bibinfo {author} {\bibfnamefont {A.}~\bibnamefont
  {Katselas}}, \bibinfo {author} {\bibfnamefont {R.}~\bibnamefont {Parin}}, \
  and\ \bibinfo {author} {\bibfnamefont {C.}~\bibnamefont {Neto}},\ }\href
  {\doibase 10.1002/admi.202200246} {\bibfield  {journal} {\bibinfo  {journal}
  {Adv. Mater. Interfaces}\ }\textbf {\bibinfo {volume} {9}},\ \bibinfo {pages}
  {1} (\bibinfo {year} {2022})}\BibitemShut {NoStop}%
\bibitem [{\citenamefont {{RN. Leach}}\ \emph {et~al.}(2006)\citenamefont {{RN.
  Leach}}, \citenamefont {Stevens}, \citenamefont {{SC. Langford}},\ and\
  \citenamefont {{JT. Dickinson}}}]{Leach2006}%
  \BibitemOpen
  \bibfield  {author} {\bibinfo {author} {\bibnamefont {{RN. Leach}}}, \bibinfo
  {author} {\bibfnamefont {F.}~\bibnamefont {Stevens}}, \bibinfo {author}
  {\bibnamefont {{SC. Langford}}}, \ and\ \bibinfo {author} {\bibnamefont {{JT.
  Dickinson}}},\ }\href {\doibase 10.1021/la061901+} {\bibfield  {journal}
  {\bibinfo  {journal} {Langmuir}\ }\textbf {\bibinfo {volume} {22}},\ \bibinfo
  {pages} {8864} (\bibinfo {year} {2006})}\BibitemShut {NoStop}%
\bibitem [{\citenamefont {Phadnis}\ and\ \citenamefont
  {Rykaczewski}(2017)}]{Phadnis2017}%
  \BibitemOpen
  \bibfield  {author} {\bibinfo {author} {\bibfnamefont {A.}~\bibnamefont
  {Phadnis}}\ and\ \bibinfo {author} {\bibfnamefont {K.}~\bibnamefont
  {Rykaczewski}},\ }\href {\doibase 10.1021/acs.langmuir.7b03141} {\bibfield
  {journal} {\bibinfo  {journal} {Langmuir}\ }\textbf {\bibinfo {volume}
  {33}},\ \bibinfo {pages} {12095} (\bibinfo {year} {2017})}\BibitemShut
  {NoStop}%
\bibitem [{\citenamefont {Sokuler}\ \emph
  {et~al.}(2010{\natexlab{a}})\citenamefont {Sokuler}, \citenamefont {{GK.
  Auernhammer}}, \citenamefont {Roth}, \citenamefont {Liu}, \citenamefont
  {Bonacurrso},\ and\ \citenamefont {{HJ. Butt}}}]{Sokuler2010}%
  \BibitemOpen
  \bibfield  {author} {\bibinfo {author} {\bibfnamefont {M.}~\bibnamefont
  {Sokuler}}, \bibinfo {author} {\bibnamefont {{GK. Auernhammer}}}, \bibinfo
  {author} {\bibfnamefont {M.}~\bibnamefont {Roth}}, \bibinfo {author}
  {\bibfnamefont {C.}~\bibnamefont {Liu}}, \bibinfo {author} {\bibfnamefont
  {E.}~\bibnamefont {Bonacurrso}}, \ and\ \bibinfo {author} {\bibnamefont {{HJ.
  Butt}}},\ }\href {\doibase 10.1021/la903996j} {\bibfield  {journal} {\bibinfo
   {journal} {Langmuir}\ }\textbf {\bibinfo {volume} {26}},\ \bibinfo {pages}
  {1544} (\bibinfo {year} {2010}{\natexlab{a}})}\BibitemShut {NoStop}%
\bibitem [{\citenamefont {Stricker}\ \emph {et~al.}(2022)\citenamefont
  {Stricker}, \citenamefont {Grillo}, \citenamefont {{EA. Marquez}.},
  \citenamefont {Panzarasa}, \citenamefont {Smith-Mannschott},\ and\
  \citenamefont {Vollmer}}]{Stricker2022universality}%
  \BibitemOpen
  \bibfield  {author} {\bibinfo {author} {\bibfnamefont {L.}~\bibnamefont
  {Stricker}}, \bibinfo {author} {\bibfnamefont {F.}~\bibnamefont {Grillo}},
  \bibinfo {author} {\bibnamefont {{EA. Marquez}.}}, \bibinfo {author}
  {\bibfnamefont {G.}~\bibnamefont {Panzarasa}}, \bibinfo {author}
  {\bibfnamefont {K.}~\bibnamefont {Smith-Mannschott}}, \ and\ \bibinfo
  {author} {\bibfnamefont {J.}~\bibnamefont {Vollmer}},\ }\href {\doibase
  10.1103/PhysRevResearch.4.L012019} {\bibfield  {journal} {\bibinfo  {journal}
  {Phys. Rev. Res.}\ }\textbf {\bibinfo {volume} {4}},\ \bibinfo {pages}
  {L012019} (\bibinfo {year} {2022})}\BibitemShut {NoStop}%
\bibitem [{\citenamefont {Anand}\ \emph {et~al.}(2015)\citenamefont {Anand},
  \citenamefont {Rykaczewski}, \citenamefont {{SB. Subramanyam}}, \citenamefont
  {Beysens},\ and\ \citenamefont {{KK. Varanasi}}}]{Anand2015}%
  \BibitemOpen
  \bibfield  {author} {\bibinfo {author} {\bibfnamefont {S.}~\bibnamefont
  {Anand}}, \bibinfo {author} {\bibfnamefont {K.}~\bibnamefont {Rykaczewski}},
  \bibinfo {author} {\bibnamefont {{SB. Subramanyam}}}, \bibinfo {author}
  {\bibfnamefont {D.}~\bibnamefont {Beysens}}, \ and\ \bibinfo {author}
  {\bibnamefont {{KK. Varanasi}}},\ }\href {\doibase 10.1039/c4sm01424c}
  {\bibfield  {journal} {\bibinfo  {journal} {Soft Matter}\ }\textbf {\bibinfo
  {volume} {11}},\ \bibinfo {pages} {69} (\bibinfo {year} {2015})}\BibitemShut
  {NoStop}%
\bibitem [{\citenamefont {{CS. Sharma}}\ \emph {et~al.}(2022)\citenamefont
  {{CS. Sharma}}, \citenamefont {Milionis}, \citenamefont {Naga}, \citenamefont
  {{CWE. Lam}}, \citenamefont {Rodriguez}, \citenamefont {{Del Ponte}},
  \citenamefont {Negri}, \citenamefont {Raoul}, \citenamefont {D'Acunzi},
  \citenamefont {{HJ. Butt}}, \citenamefont {Vollmer},\ and\ \citenamefont
  {Poulikakos}}]{Sharma2022}%
  \BibitemOpen
  \bibfield  {author} {\bibinfo {author} {\bibnamefont {{CS. Sharma}}},
  \bibinfo {author} {\bibfnamefont {A.}~\bibnamefont {Milionis}}, \bibinfo
  {author} {\bibfnamefont {A.}~\bibnamefont {Naga}}, \bibinfo {author}
  {\bibnamefont {{CWE. Lam}}}, \bibinfo {author} {\bibfnamefont
  {G.}~\bibnamefont {Rodriguez}}, \bibinfo {author} {\bibfnamefont {M.~F.}\
  \bibnamefont {{Del Ponte}}}, \bibinfo {author} {\bibfnamefont
  {V.}~\bibnamefont {Negri}}, \bibinfo {author} {\bibfnamefont
  {H.}~\bibnamefont {Raoul}}, \bibinfo {author} {\bibfnamefont
  {M.}~\bibnamefont {D'Acunzi}}, \bibinfo {author} {\bibnamefont {{HJ. Butt}}},
  \bibinfo {author} {\bibfnamefont {D.}~\bibnamefont {Vollmer}}, \ and\
  \bibinfo {author} {\bibfnamefont {D.}~\bibnamefont {Poulikakos}},\ }\href
  {\doibase 10.1002/adfm.202109633} {\bibfield  {journal} {\bibinfo  {journal}
  {Adv. Funct. Mater.}\ }\textbf {\bibinfo {volume} {32}} (\bibinfo {year}
  {2022}),\ 10.1002/adfm.202109633}\BibitemShut {NoStop}%
\bibitem [{\citenamefont {Steyer}\ \emph {et~al.}(1990)\citenamefont {Steyer},
  \citenamefont {Guenoun}, \citenamefont {Beysens},\ and\ \citenamefont {{CM.
  Knobler}}}]{Steyer1990}%
  \BibitemOpen
  \bibfield  {author} {\bibinfo {author} {\bibfnamefont {A.}~\bibnamefont
  {Steyer}}, \bibinfo {author} {\bibfnamefont {P.}~\bibnamefont {Guenoun}},
  \bibinfo {author} {\bibfnamefont {D.}~\bibnamefont {Beysens}}, \ and\
  \bibinfo {author} {\bibnamefont {{CM. Knobler}}},\ }\href {\doibase
  10.1103/PhysRevB.42.1086} {\bibfield  {journal} {\bibinfo  {journal} {Phys.
  Rev. B}\ }\textbf {\bibinfo {volume} {42}},\ \bibinfo {pages} {1086}
  (\bibinfo {year} {1990})}\BibitemShut {NoStop}%
\bibitem [{\citenamefont {Steyer}\ \emph {et~al.}(1993)\citenamefont {Steyer},
  \citenamefont {Guenoun},\ and\ \citenamefont {Beysens}}]{Steyer1993}%
  \BibitemOpen
  \bibfield  {author} {\bibinfo {author} {\bibfnamefont {A.}~\bibnamefont
  {Steyer}}, \bibinfo {author} {\bibfnamefont {P.}~\bibnamefont {Guenoun}}, \
  and\ \bibinfo {author} {\bibfnamefont {D.}~\bibnamefont {Beysens}},\ }\href
  {\doibase 10.1103/PhysRevE.48.428} {\bibfield  {journal} {\bibinfo  {journal}
  {Phys. Rev. E}\ }\textbf {\bibinfo {volume} {48}},\ \bibinfo {pages} {428}
  (\bibinfo {year} {1993})}\BibitemShut {NoStop}%
\bibitem [{\citenamefont {Beysens}\ and\ \citenamefont {{CM.
  Knobler}}(1986)}]{beysens1986growth}%
  \BibitemOpen
  \bibfield  {author} {\bibinfo {author} {\bibfnamefont {D.}~\bibnamefont
  {Beysens}}\ and\ \bibinfo {author} {\bibnamefont {{CM. Knobler}}},\
  }\href@noop {} {\bibfield  {journal} {\bibinfo  {journal} {Phys. Rev. Lett.}\
  }\textbf {\bibinfo {volume} {57}},\ \bibinfo {pages} {1433} (\bibinfo {year}
  {1986})}\BibitemShut {NoStop}%
\bibitem [{\citenamefont {{JL. Viovy}}\ \emph {et~al.}(1988)\citenamefont {{JL.
  Viovy}}, \citenamefont {Beysens},\ and\ \citenamefont {{CM.
  Knobler}}}]{Viovy1988scaling}%
  \BibitemOpen
  \bibfield  {author} {\bibinfo {author} {\bibnamefont {{JL. Viovy}}}, \bibinfo
  {author} {\bibfnamefont {D.}~\bibnamefont {Beysens}}, \ and\ \bibinfo
  {author} {\bibnamefont {{CM. Knobler}}},\ }\href@noop {} {\bibfield
  {journal} {\bibinfo  {journal} {Phys. Rev. A}\ }\textbf {\bibinfo {volume}
  {37}},\ \bibinfo {pages} {4965} (\bibinfo {year} {1988})}\BibitemShut
  {NoStop}%
\bibitem [{\citenamefont {Family}\ and\ \citenamefont
  {Meakin}(1988)}]{family1988scaling}%
  \BibitemOpen
  \bibfield  {author} {\bibinfo {author} {\bibfnamefont {F.}~\bibnamefont
  {Family}}\ and\ \bibinfo {author} {\bibfnamefont {P.}~\bibnamefont
  {Meakin}},\ }\href@noop {} {\bibfield  {journal} {\bibinfo  {journal} {Phys.
  Rev. Lett.}\ }\textbf {\bibinfo {volume} {61}},\ \bibinfo {pages} {428}
  (\bibinfo {year} {1988})}\BibitemShut {NoStop}%
\bibitem [{\citenamefont {Kolb}(1989)}]{Kolb1989}%
  \BibitemOpen
  \bibfield  {author} {\bibinfo {author} {\bibfnamefont {M.}~\bibnamefont
  {Kolb}},\ }\href {\doibase 10.1103/PhysRevLett.62.1699} {\bibfield  {journal}
  {\bibinfo  {journal} {Phys. Rev. Lett.}\ }\textbf {\bibinfo {volume} {62}},\
  \bibinfo {pages} {1699} (\bibinfo {year} {1989})}\BibitemShut {NoStop}%
\bibitem [{\citenamefont {Blaschke}\ \emph {et~al.}(2012)\citenamefont
  {Blaschke}, \citenamefont {Lapp}, \citenamefont {Hof},\ and\ \citenamefont
  {Vollmer}}]{blaschke2012breath}%
  \BibitemOpen
  \bibfield  {author} {\bibinfo {author} {\bibfnamefont {J.}~\bibnamefont
  {Blaschke}}, \bibinfo {author} {\bibfnamefont {T.}~\bibnamefont {Lapp}},
  \bibinfo {author} {\bibfnamefont {B.}~\bibnamefont {Hof}}, \ and\ \bibinfo
  {author} {\bibfnamefont {J.}~\bibnamefont {Vollmer}},\ }\href@noop {}
  {\bibfield  {journal} {\bibinfo  {journal} {Phys. Rev. Lett.}\ }\textbf
  {\bibinfo {volume} {109}},\ \bibinfo {pages} {068701} (\bibinfo {year}
  {2012})}\BibitemShut {NoStop}%
\bibitem [{\citenamefont {{NV. Brilliantov}}\ \emph {et~al.}(1998)\citenamefont
  {{NV. Brilliantov}}, \citenamefont {Andrienko}, \citenamefont {{PL.
  Krapivsky}},\ and\ \citenamefont {Kurths}}]{brilliantov1998polydisperse}%
  \BibitemOpen
  \bibfield  {author} {\bibinfo {author} {\bibnamefont {{NV. Brilliantov}}},
  \bibinfo {author} {\bibfnamefont {Y.~A.}\ \bibnamefont {Andrienko}}, \bibinfo
  {author} {\bibnamefont {{PL. Krapivsky}}}, \ and\ \bibinfo {author}
  {\bibfnamefont {J.}~\bibnamefont {Kurths}},\ }\href@noop {} {\bibfield
  {journal} {\bibinfo  {journal} {Phys. Rev. E}\ }\textbf {\bibinfo {volume}
  {58}},\ \bibinfo {pages} {3530} (\bibinfo {year} {1998})}\BibitemShut
  {NoStop}%
\bibitem [{\citenamefont {Family}\ and\ \citenamefont
  {Meakin}(1989)}]{family1989kinetics}%
  \BibitemOpen
  \bibfield  {author} {\bibinfo {author} {\bibfnamefont {F.}~\bibnamefont
  {Family}}\ and\ \bibinfo {author} {\bibfnamefont {P.}~\bibnamefont
  {Meakin}},\ }\href@noop {} {\bibfield  {journal} {\bibinfo  {journal} {Phys.
  Rev. A}\ }\textbf {\bibinfo {volume} {40}},\ \bibinfo {pages} {3836}
  (\bibinfo {year} {1989})}\BibitemShut {NoStop}%
\bibitem [{\citenamefont {{IF. Guha}}\ \emph {et~al.}(2017)\citenamefont {{IF.
  Guha}}, \citenamefont {Anand},\ and\ \citenamefont {{KK.
  Varanasi}.}}]{Guha2017}%
  \BibitemOpen
  \bibfield  {author} {\bibinfo {author} {\bibnamefont {{IF. Guha}}}, \bibinfo
  {author} {\bibfnamefont {S.}~\bibnamefont {Anand}}, \ and\ \bibinfo {author}
  {\bibnamefont {{KK. Varanasi}.}},\ }\href {\doibase
  10.1038/s41467-017-01420-8} {\bibfield  {journal} {\bibinfo  {journal} {Nat.
  Commun.}\ }\textbf {\bibinfo {volume} {8}},\ \bibinfo {pages} {1} (\bibinfo
  {year} {2017})}\BibitemShut {NoStop}%
\bibitem [{\citenamefont {Baratian}\ \emph {et~al.}(2018)\citenamefont
  {Baratian}, \citenamefont {Dey}, \citenamefont {Hoek}, \citenamefont {{D. Van
  Den Ende}},\ and\ \citenamefont {Mugele}}]{Baratian2018}%
  \BibitemOpen
  \bibfield  {author} {\bibinfo {author} {\bibfnamefont {D.}~\bibnamefont
  {Baratian}}, \bibinfo {author} {\bibfnamefont {R.}~\bibnamefont {Dey}},
  \bibinfo {author} {\bibfnamefont {H.}~\bibnamefont {Hoek}}, \bibinfo {author}
  {\bibnamefont {{D. Van Den Ende}}}, \ and\ \bibinfo {author} {\bibfnamefont
  {F.}~\bibnamefont {Mugele}},\ }\href {\doibase
  10.1103/PhysRevLett.120.214502} {\bibfield  {journal} {\bibinfo  {journal}
  {Phys. Rev. Lett.}\ }\textbf {\bibinfo {volume} {120}},\ \bibinfo {pages}
  {214502} (\bibinfo {year} {2018})}\BibitemShut {NoStop}%
\bibitem [{\citenamefont {B{\"o}ker}\ \emph {et~al.}(2004)\citenamefont
  {B{\"o}ker}, \citenamefont {Lin}, \citenamefont {Chiapperini}, \citenamefont
  {Horowitz}, \citenamefont {Thompson}, \citenamefont {Carreon}, \citenamefont
  {Xu}, \citenamefont {Abetz}, \citenamefont {Skaff}, \citenamefont {{AD.
  Dinsmore}}, \citenamefont {Emrick},\ and\ \citenamefont {{TP.
  Russell}}}]{boker2004hierarchical}%
  \BibitemOpen
  \bibfield  {author} {\bibinfo {author} {\bibfnamefont {A.}~\bibnamefont
  {B{\"o}ker}}, \bibinfo {author} {\bibfnamefont {Y.}~\bibnamefont {Lin}},
  \bibinfo {author} {\bibfnamefont {K.}~\bibnamefont {Chiapperini}}, \bibinfo
  {author} {\bibfnamefont {R.}~\bibnamefont {Horowitz}}, \bibinfo {author}
  {\bibfnamefont {M.}~\bibnamefont {Thompson}}, \bibinfo {author}
  {\bibfnamefont {V.}~\bibnamefont {Carreon}}, \bibinfo {author} {\bibfnamefont
  {T.}~\bibnamefont {Xu}}, \bibinfo {author} {\bibfnamefont {C.}~\bibnamefont
  {Abetz}}, \bibinfo {author} {\bibfnamefont {H.}~\bibnamefont {Skaff}},
  \bibinfo {author} {\bibnamefont {{AD. Dinsmore}}}, \bibinfo {author}
  {\bibfnamefont {T.}~\bibnamefont {Emrick}}, \ and\ \bibinfo {author}
  {\bibnamefont {{TP. Russell}}},\ }\href@noop {} {\bibfield  {journal}
  {\bibinfo  {journal} {Nat. Mater.}\ }\textbf {\bibinfo {volume} {3}},\
  \bibinfo {pages} {302} (\bibinfo {year} {2004})}\BibitemShut {NoStop}%
\bibitem [{\citenamefont {Zhang}\ \emph {et~al.}(2015)\citenamefont {Zhang},
  \citenamefont {Bai},\ and\ \citenamefont {Li}}]{zhang2015breath}%
  \BibitemOpen
  \bibfield  {author} {\bibinfo {author} {\bibfnamefont {A.}~\bibnamefont
  {Zhang}}, \bibinfo {author} {\bibfnamefont {H.}~\bibnamefont {Bai}}, \ and\
  \bibinfo {author} {\bibfnamefont {L.}~\bibnamefont {Li}},\ }\href@noop {}
  {\bibfield  {journal} {\bibinfo  {journal} {Chem. Rev.}\ }\textbf {\bibinfo
  {volume} {115}},\ \bibinfo {pages} {9801} (\bibinfo {year}
  {2015})}\BibitemShut {NoStop}%
\bibitem [{\citenamefont {{RW. Style}}\ \emph {et~al.}(2013)\citenamefont {{RW.
  Style}}, \citenamefont {Che}, \citenamefont {{SJ. Park}}, \citenamefont {{BM.
  Weon}}, \citenamefont {{JH. Je}}, \citenamefont {Hyland}, \citenamefont {{GK.
  German}}, \citenamefont {{MP. Power}}, \citenamefont {{LA. Wilen}},
  \citenamefont {{JS. Wettlaufer}},\ and\ \citenamefont {{ER.
  Dufresne}}}]{style2013patterning}%
  \BibitemOpen
  \bibfield  {author} {\bibinfo {author} {\bibnamefont {{RW. Style}}}, \bibinfo
  {author} {\bibfnamefont {Y.}~\bibnamefont {Che}}, \bibinfo {author}
  {\bibnamefont {{SJ. Park}}}, \bibinfo {author} {\bibnamefont {{BM. Weon}}},
  \bibinfo {author} {\bibnamefont {{JH. Je}}}, \bibinfo {author} {\bibfnamefont
  {C.}~\bibnamefont {Hyland}}, \bibinfo {author} {\bibnamefont {{GK. German}}},
  \bibinfo {author} {\bibnamefont {{MP. Power}}}, \bibinfo {author}
  {\bibnamefont {{LA. Wilen}}}, \bibinfo {author} {\bibnamefont {{JS.
  Wettlaufer}}}, \ and\ \bibinfo {author} {\bibnamefont {{ER. Dufresne}}},\
  }\href {\doibase 10.1073/pnas.1307122110} {\bibfield  {journal} {\bibinfo
  {journal} {Proc. Natl. Acad. Sci.}\ }\textbf {\bibinfo {volume} {110}},\
  \bibinfo {pages} {12541} (\bibinfo {year} {2013})}\BibitemShut {NoStop}%
\bibitem [{\citenamefont {{FJ. Dent}}\ \emph {et~al.}(2022)\citenamefont {{FJ.
  Dent}}, \citenamefont {Harbottle}, \citenamefont {{NJ. Warren}},\ and\
  \citenamefont {Khodaparast}}]{dent2022temporally}%
  \BibitemOpen
  \bibfield  {author} {\bibinfo {author} {\bibnamefont {{FJ. Dent}}}, \bibinfo
  {author} {\bibfnamefont {D.}~\bibnamefont {Harbottle}}, \bibinfo {author}
  {\bibnamefont {{NJ. Warren}}}, \ and\ \bibinfo {author} {\bibfnamefont
  {S.}~\bibnamefont {Khodaparast}},\ }\href {\doibase 10.1021/acsami.2c05635}
  {\bibfield  {journal} {\bibinfo  {journal} {ACS Appl. Mater. Interfaces}\
  }\textbf {\bibinfo {volume} {14}},\ \bibinfo {pages} {27435} (\bibinfo {year}
  {2022})}\BibitemShut {NoStop}%
\bibitem [{\citenamefont {Andreotti}\ and\ \citenamefont {{JH.
  Snoeijer}}(2020)}]{Andreotti2020Annual}%
  \BibitemOpen
  \bibfield  {author} {\bibinfo {author} {\bibfnamefont {B.}~\bibnamefont
  {Andreotti}}\ and\ \bibinfo {author} {\bibnamefont {{JH. Snoeijer}}},\ }\href
  {\doibase 10.1146/annurev-fluid-010719-060147} {\bibfield  {journal}
  {\bibinfo  {journal} {Annu. Rev. Fluid Mech.}\ }\textbf {\bibinfo {volume}
  {52}},\ \bibinfo {pages} {285} (\bibinfo {year} {2020})}\BibitemShut
  {NoStop}%
\bibitem [{\citenamefont {Karpitschka}\ \emph {et~al.}(2016)\citenamefont
  {Karpitschka}, \citenamefont {Pandey}, \citenamefont {{LA. Lubbers}},
  \citenamefont {{JH. Weijs}}, \citenamefont {Botto}, \citenamefont {Das},
  \citenamefont {Andreotti},\ and\ \citenamefont {{JH.
  Snoeijer}}}]{Karpitschka2016}%
  \BibitemOpen
  \bibfield  {author} {\bibinfo {author} {\bibfnamefont {S.}~\bibnamefont
  {Karpitschka}}, \bibinfo {author} {\bibfnamefont {A.}~\bibnamefont {Pandey}},
  \bibinfo {author} {\bibnamefont {{LA. Lubbers}}}, \bibinfo {author}
  {\bibnamefont {{JH. Weijs}}}, \bibinfo {author} {\bibfnamefont
  {L.}~\bibnamefont {Botto}}, \bibinfo {author} {\bibfnamefont
  {S.}~\bibnamefont {Das}}, \bibinfo {author} {\bibfnamefont {B.}~\bibnamefont
  {Andreotti}}, \ and\ \bibinfo {author} {\bibnamefont {{JH. Snoeijer}}},\
  }\href {\doibase 10.1073/pnas.1601411113} {\bibfield  {journal} {\bibinfo
  {journal} {Proc. Natl. Acad. Sci.}\ }\textbf {\bibinfo {volume} {113}},\
  \bibinfo {pages} {7403} (\bibinfo {year} {2016})}\BibitemShut {NoStop}%
\bibitem [{\citenamefont {Sokuler}\ \emph
  {et~al.}(2010{\natexlab{b}})\citenamefont {Sokuler}, \citenamefont {{GK.
  Auernhammer}}, \citenamefont {{CJ. Liu}}, \citenamefont {Bonaccurso},\ and\
  \citenamefont {{HJ. Butt}}}]{Sokuler2010interspacing}%
  \BibitemOpen
  \bibfield  {author} {\bibinfo {author} {\bibfnamefont {M.}~\bibnamefont
  {Sokuler}}, \bibinfo {author} {\bibnamefont {{GK. Auernhammer}}}, \bibinfo
  {author} {\bibnamefont {{CJ. Liu}}}, \bibinfo {author} {\bibfnamefont
  {E.}~\bibnamefont {Bonaccurso}}, \ and\ \bibinfo {author} {\bibnamefont {{HJ.
  Butt}}},\ }\href {\doibase 10.1209/0295-5075/89/36004} {\bibfield  {journal}
  {\bibinfo  {journal} {EPL}\ }\textbf {\bibinfo {volume} {89}},\ \bibinfo
  {pages} {36004} (\bibinfo {year} {2010}{\natexlab{b}})}\BibitemShut {NoStop}%
\bibitem [{\citenamefont {Bouillant}\ \emph {et~al.}(2024)\citenamefont
  {Bouillant}, \citenamefont {{JH. Snoeijer}},\ and\ \citenamefont
  {Andreotti}}]{paper_bis}%
  \BibitemOpen
  \bibfield  {author} {\bibinfo {author} {\bibfnamefont {A.}~\bibnamefont
  {Bouillant}}, \bibinfo {author} {\bibnamefont {{JH. Snoeijer}}}, \ and\
  \bibinfo {author} {\bibfnamefont {B.}~\bibnamefont {Andreotti}},\ }\href@noop
  {} {\bibfield  {journal} {\bibinfo  {journal} {Submitted as a companion
  paper}\ } (\bibinfo {year} {2024})}\BibitemShut {NoStop}%
\bibitem [{\citenamefont {{YO. Popov}}(2005)}]{Popov2005}%
  \BibitemOpen
  \bibfield  {author} {\bibinfo {author} {\bibnamefont {{YO. Popov}}},\ }\href
  {\doibase 10.1103/PhysRevE.71.036313} {\bibfield  {journal} {\bibinfo
  {journal} {Phys. Rev. E}\ }\textbf {\bibinfo {volume} {71}},\ \bibinfo
  {pages} {036313} (\bibinfo {year} {2005})}\BibitemShut {NoStop}%
\bibitem [{\citenamefont {{JW. Rose}}(2002)}]{Rose2002}%
  \BibitemOpen
  \bibfield  {author} {\bibinfo {author} {\bibnamefont {{JW. Rose}}},\ }\href
  {\doibase 10.1243/09576500260049034} {\bibfield  {journal} {\bibinfo
  {journal} {Proc. Inst. Mech. Eng. Part A J. Power Energy}\ }\textbf {\bibinfo
  {volume} {216}},\ \bibinfo {pages} {115} (\bibinfo {year}
  {2002})}\BibitemShut {NoStop}%
\bibitem [{\citenamefont {{BS, Sikarwar}}\ \emph {et~al.}(2012)\citenamefont
  {{BS, Sikarwar}}, \citenamefont {Khandekar}, \citenamefont {Agrawal},
  \citenamefont {Kumar},\ and\ \citenamefont {Muralidhar}}]{Sikarwar2012}%
  \BibitemOpen
  \bibfield  {author} {\bibinfo {author} {\bibnamefont {{BS, Sikarwar}}},
  \bibinfo {author} {\bibfnamefont {S.}~\bibnamefont {Khandekar}}, \bibinfo
  {author} {\bibfnamefont {S.}~\bibnamefont {Agrawal}}, \bibinfo {author}
  {\bibfnamefont {S.}~\bibnamefont {Kumar}}, \ and\ \bibinfo {author}
  {\bibfnamefont {K.}~\bibnamefont {Muralidhar}},\ }\href {\doibase
  10.1080/01457632.2012.611463} {\bibfield  {journal} {\bibinfo  {journal}
  {Heat Transf. Eng.}\ }\textbf {\bibinfo {volume} {33}},\ \bibinfo {pages}
  {301} (\bibinfo {year} {2012})}\BibitemShut {NoStop}%
\bibitem [{\citenamefont {Zhao}\ and\ \citenamefont
  {Beysens}(1995)}]{Zhao1995}%
  \BibitemOpen
  \bibfield  {author} {\bibinfo {author} {\bibfnamefont {H.}~\bibnamefont
  {Zhao}}\ and\ \bibinfo {author} {\bibfnamefont {D.}~\bibnamefont {Beysens}},\
  }\href {\doibase 10.1021/la00002a045} {\bibfield  {journal} {\bibinfo
  {journal} {Langmuir}\ }\textbf {\bibinfo {volume} {11}},\ \bibinfo {pages}
  {627} (\bibinfo {year} {1995})}\BibitemShut {NoStop}%
\bibitem [{\citenamefont {Nepomnyashchy}\ \emph {et~al.}(2006)\citenamefont
  {Nepomnyashchy}, \citenamefont {Golovin}, \citenamefont {Tikhomirova},\ and\
  \citenamefont {Volpert}}]{Nepomnyashchy2006}%
  \BibitemOpen
  \bibfield  {author} {\bibinfo {author} {\bibfnamefont {A.}~\bibnamefont
  {Nepomnyashchy}}, \bibinfo {author} {\bibfnamefont {A.}~\bibnamefont
  {Golovin}}, \bibinfo {author} {\bibfnamefont {A.}~\bibnamefont
  {Tikhomirova}}, \ and\ \bibinfo {author} {\bibfnamefont {V.}~\bibnamefont
  {Volpert}},\ }\href {\doibase 10.1103/PhysRevE.74.021605} {\bibfield
  {journal} {\bibinfo  {journal} {Phys. Rev. E}\ }\textbf {\bibinfo {volume}
  {74}},\ \bibinfo {pages} {1} (\bibinfo {year} {2006})}\BibitemShut {NoStop}%
\bibitem [{\citenamefont {{CM. Knobler}}\ and\ \citenamefont
  {Beysens}(1988)}]{Knobler1988}%
  \BibitemOpen
  \bibfield  {author} {\bibinfo {author} {\bibnamefont {{CM. Knobler}}}\ and\
  \bibinfo {author} {\bibfnamefont {D.}~\bibnamefont {Beysens}},\ }\href
  {\doibase 10.1209/0295-5075/6/8/007} {\bibfield  {journal} {\bibinfo
  {journal} {EPL}\ }\textbf {\bibinfo {volume} {6}},\ \bibinfo {pages} {707}
  (\bibinfo {year} {1988})}\BibitemShut {NoStop}%
\bibitem [{\citenamefont {Zhang}\ \emph {et~al.}(2020)\citenamefont {Zhang},
  \citenamefont {{RA. Mei}}, \citenamefont {Botto},\ and\ \citenamefont
  {Yang}}]{Zhang2020}%
  \BibitemOpen
  \bibfield  {author} {\bibinfo {author} {\bibfnamefont {R.}~\bibnamefont
  {Zhang}}, \bibinfo {author} {\bibnamefont {{RA. Mei}}}, \bibinfo {author}
  {\bibfnamefont {L.}~\bibnamefont {Botto}}, \ and\ \bibinfo {author}
  {\bibfnamefont {Z.}~\bibnamefont {Yang}},\ }\href {\doibase
  10.1021/acs.langmuir.9b03806} {\bibfield  {journal} {\bibinfo  {journal}
  {Langmuir}\ }\textbf {\bibinfo {volume} {36}},\ \bibinfo {pages} {5400}
  (\bibinfo {year} {2020})}\BibitemShut {NoStop}%
\bibitem [{\citenamefont {Merikanto}\ \emph {et~al.}(2007)\citenamefont
  {Merikanto}, \citenamefont {Zapadinsky}, \citenamefont {Lauri},\ and\
  \citenamefont {Vehkam\"aki}}]{Merikanto2007}%
  \BibitemOpen
  \bibfield  {author} {\bibinfo {author} {\bibfnamefont {J.}~\bibnamefont
  {Merikanto}}, \bibinfo {author} {\bibfnamefont {E.}~\bibnamefont
  {Zapadinsky}}, \bibinfo {author} {\bibfnamefont {A.}~\bibnamefont {Lauri}}, \
  and\ \bibinfo {author} {\bibfnamefont {H.}~\bibnamefont {Vehkam\"aki}},\
  }\href {\doibase 10.1103/PhysRevLett.98.145702} {\bibfield  {journal}
  {\bibinfo  {journal} {Phys. Rev. Lett.}\ }\textbf {\bibinfo {volume} {98}},\
  \bibinfo {pages} {145702} (\bibinfo {year} {2007})}\BibitemShut {NoStop}%
\bibitem [{\citenamefont {Li}\ and\ \citenamefont
  {Signorell}(2021)}]{li2021understanding}%
  \BibitemOpen
  \bibfield  {author} {\bibinfo {author} {\bibfnamefont {C.}~\bibnamefont
  {Li}}\ and\ \bibinfo {author} {\bibfnamefont {R.}~\bibnamefont {Signorell}},\
  }\href {\doibase 10.1016/j.jaerosci.2020.105676} {\bibfield  {journal}
  {\bibinfo  {journal} {J. Aerosol Sci.}\ }\textbf {\bibinfo {volume} {153}},\
  \bibinfo {pages} {105676} (\bibinfo {year} {2021})}\BibitemShut {NoStop}%
\end{thebibliography}%

\end{document}